\documentstyle[11pt,blois_proc,epsfig]{article}

\def\be{\begin{equation}}
\def\ee{\end{equation}}
\def\bea{\begin{eqnarray}}
\def\eea{\end{eqnarray}}

\newcommand{\GeVc}    {\mbox{$ {\mathrm{GeV}}/c                            $}}
\newcommand{\GeVcc}    {\mbox{$ {\mathrm{GeV}}/c^2                           $}}
\newcommand{\MeVc}    {\mbox{$ {\mathrm{MeV}}/c                            $}}

\begin{document}
\vspace{-3.cm}
\noindent
XIth Rencontres de Blois (Frontiers of Matter) proceedings.\\
ISN-99.98\\
\vspace*{2.cm}
\title{PERFORMANCE RESULTS OF THE AMS-01 AEROGEL THRESHOLD \Cher}

\author{ F. MAYET \\ (for the AMS collaboration)}

\address{Institut des Sciences Nucl\'eaires, 53 avenue des Martyrs, 38026 Grenoble cedex, France\\
E-mail: Frederic.Mayet@isn.in2p3.fr}

\maketitle\abstracts{
The Alpha Magnetic Spectrometer (AMS) was flown in june 1998 on board of the space shuttle Discovery (flight STS-91) 
at an altitude ranging between 320 and 390 km.
This preliminary version of AMS included an Aerogel Threshold \cher detector (ATC) to separate 
$\bar{p}$ from $e^{-}$ background, for momenta less than 3.5 \GeVc.
In this paper, the design and physical principles of ATC will be discussed briefly, then the performance results of
ATC will be presented.}

\section{Role of ATC in AMS-01}
One of the purposes of the AMS Shuttle flight was to search for cosmic $\bar{p}$ and to measure 
${\bar{p}}/p$ ratio for momenta(P) below  3.5~\GeVc. The major background component to the $\bar{p}$ sample is expected to be $e^{-}$ (${\bar{p}}/e^{-} \sim 10^{-2}$ for the considered
P range).\\
Using AMS-01 momentum resolution~\cite{bill} (${\Delta P}/P=7.\%$) and $\beta$ resolution~\cite{chou} (${\Delta \beta }/\beta \simeq
3.3\%$), it can be shown~\cite{atc} that
AMS electron rejection falls sharply above 1.5-2 \GeVc~. ATC is used to extend $\bar{p}/e^{-}$ discrimination range up to 3.5
\GeVc. As a secondary result, ${p}/{e^{+}}$ separation can also be improved using appropriate ATC selections.

\noindent
Many balloon-borne experiments (CAPRICE, BESS,...) have included a \cher counter and a Ring Imaging \cher counter 
is under study for the next phase of the AMS experiment~\cite{thom,fer}.\\
Using an aerogel material with a low refractive index (n=1.035), 
the ATC counter profits from \cher effect to separate $\bar{p}$ from $e^{-}$ at low energy.
Indeed, the momentum threshold is 1.9 \MeVc~for $e^{\pm}$ and 3.5 \GeVc~for $p$($\bar{p}$). 
Hence, in the GeV range and up to this momentum, $p$($\bar{p}$) are not expected to produce any signal, while $e^{\pm}$ will give a full signal.
\begin{figure}
\psfig{figure=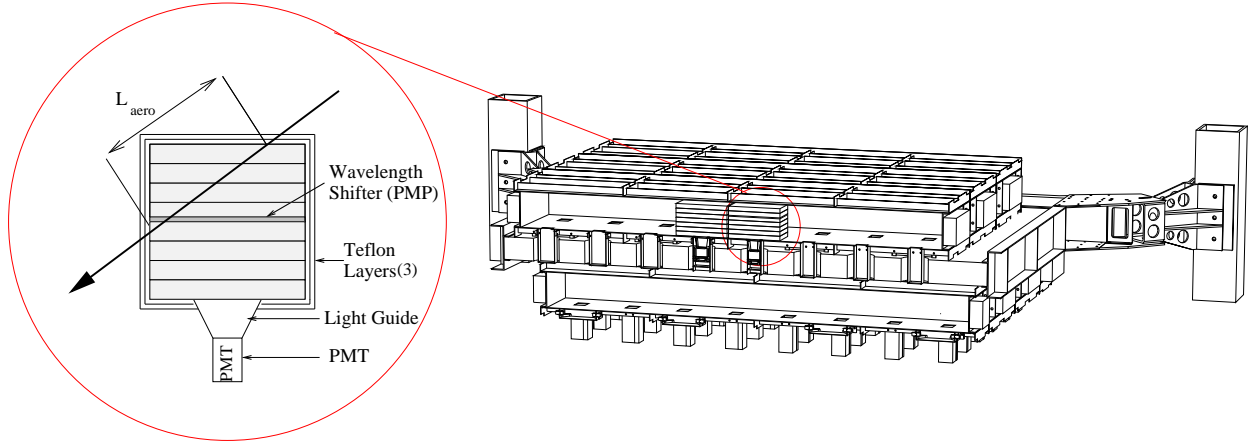,height=6.5in,angle=270}
{\noindent
\caption{ATC detector design. The 2 shifted layers (in x and y directions) can be seen together	with 
the structure allowing to mount the ATC directly on the Space Shuttle. On the detailed view of the ATC cell, 
the eight blocks of aerogel are shown together with three teflon layers and the PMP wavelength shifter in the middle of the cell.
\label{fig:atcfig}}}
\end{figure}

\noindent
The design of ATC will be described in detail elsewhere~\cite{atc}.
The elementary component of the ATC detector is the aerogel cell 
(11 $\times$ 11 $\times$ 8.8 $cm^{3}$, see figure \ref{fig:atcfig}), filled with eight $1.1 cm$ thick aerogel blocks.
The emitted photons are reflected by three $250 \mu m$ teflon layers surrounding the blocks, until they reach 
the photomultiplier's window (Hamamatsu R-5900).\\
The 168 cells are arranged in 2 layers (80 cells in the upper one and 88 cells in the lower one). 
Due to the direct photon detection used, the Rayleigh diffusion ($L_{R} \propto \lambda_{\gamma}^{4}$) and 
the absorption ($L_{abs} \propto \lambda_{\gamma}^{2}$) are known to be the most limiting processes.
These two effects are decreasing with increasing photons wavelength. For this purpose, a wavelength shifter is placed in the middle of each cell. 
It consists of a thin layer of tedlar (25 $\mu m$) soaked in a PMP solution.
This allows to shift~\footnote{It should be noticed that maximum efficiency 
of the R-5900 photomultipier tube is also at $\lambda \sim 420 nm$.} wavelength from 300$nm$ up to 420$nm$. 
The use of the shifter leads to an overall increase in number of p.e, estimated to be $\sim 40.\%$.\\

\section{ATC performance results}

More detailed results may be found in reference 3. First, ATC was not affected neither by the launch nor the space
conditions, which is a success because of the natural fragility of the aerogel material. Then, some basic checks have been done, to ensure
that the ATC response is consistent with the main physical dependences. Namely, the number of photons created by \cher effect, in a 
material of refractive index $n$, is known to be :
\begin{equation}
N_{pe} \propto L_{aero}\times Z^{2} \times (1-\frac{1}{n^{2} \beta^{2}})
\label{cherequa}  
\end{equation}
As shown in reference 3, the ATC signal as a linear dependence with the square of the particle's charge ($Z^{2}$) and the path length in the material
($L_{aero}$). Moreover, using flight data, the refractive index has been evaluated to be $n=1.034$, which is in good agreement with the known value.

\noindent
The ATC rejection and efficiency for particle selection is evaluated by defining two control samples:
Electrons (positrons) are simulated by high energy protons ($P \geq 15 $ and $\beta \geq 0.99$), 
detected near equator by AMS, thus taking advantage on the geomagnetic cutoff which ensure that most of the particles 
in this region are high energy ones.
Antiprotons are simulated by low energy protons ($P \leq 3.5$ \GeVc~, $\beta \leq 0.97$ and $0.6 \leq M \leq1.2$ \GeVcc).

\noindent
Using these two control samples, one gets 7.5 p.e for $e^{\pm}$ (after correction on various effects such as electronic
threshold and $\beta$ effect). On the other hand, most of the $p$($\bar{p}$) give 0 p.e, although ATC
encountered a residual light problem due to after-pulses 
in the PMT, \cher effect in the PMT window and scintillation effect in various materials.

\begin{figure}
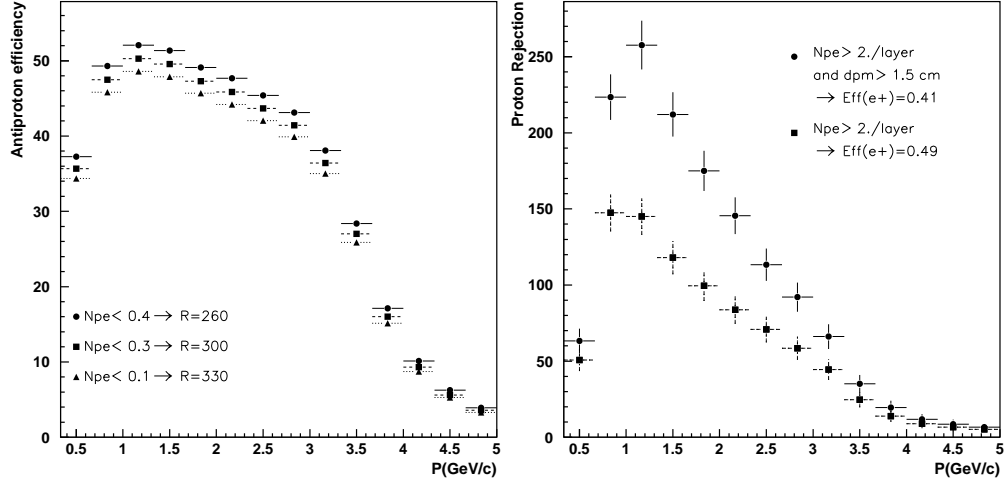

\begin{center}
\psfig{figure=effpro_bw.epsi,height=2.5in}
\psfig{figure=eplus_bw.epsi,height=2.5in}
{\noindent
\caption{Antiproton efficiency (left) as a function of P(\GeVc) for different cuts on $n_{p.e}$. 
The $e^{-}$ rejection (R) is also indicated for each cut. 
Proton rejection (right) as a function of P(\GeVc), together with $e^{+}$ efficiency($\epsilon$).
\label{fig:atcperf}}}
\end{center}
\end{figure}

\noindent
Antiprotons are selected as particles crossing two cells, leading to an overall 
geometrical efficiency of 72 $\%$, and giving less than 0.15 p.e .  
Using this selection, ATC provides a rejection of 330 against electrons, with a maximum 
efficiency of $48\%$ (shown in fig. \ref{fig:atcperf}).

\noindent
ATC may also be used to select positrons out of proton background. In this case, the ATC conditions require that the particle crosses 
2 cells and produce, in each cell, more than 2 p.e . In order to avoid contamination \footnote{For more details, see ref. 3} due to
protons passing close to the PMT, and thus producing \cher effect in the window, the closest distance 
to the PMT should be greater than 1.5 $cm$. Using this selection, ATC provides a separation between positrons and proton background, 
with an efficiency of $41 \%$ and a maximum rejection of 260 (see fig. \ref{fig:atcperf}).

\noindent
The first AMS test flight has been largely successful for AMS in general and ATC in particular. No major problems were encountered, and ATC allows 
to extend $\bar{p}/e^{-}$ discrimination range up to 3.5 \GeVc, with a good efficiency and a high rejection, as shown above.
Furthermore, ATC may be used as a redundant way of selecting $e^{+}$ out of proton background.
\section*{Acknowledgments}
The AMS Aerogel Threshold \cher is the result of the contributions of physicists, engineers and
technicians from ISN Grenoble and LAPP Annecy (France), INFN Firenze and INFN Bologna 
(Italy), ITEP Moscow (Russia), Academia Sinica (Taiwan), LIP Lisboa (Portugal).\\
The author wishes to thank Daniel Santos, Jean Favier and Fernando Barao, for the fruitful collaboration on ATC
analysis.

\end{document}